\documentclass[3p, times, procedia]{elsarticle}

\usepackage{ecrc}

\volume{00}

\firstpage{1}

\journalname{Journal of Magnetism and Magnetic Materials}

\runauth{I. V. Bychkov at al.}

\jid{jmmm}

\jnltitlelogo{Journal of Magnetism and Magnetic Materials}

\CopyrightLine{2012}{Published by Elsevier Ltd.}

\usepackage{amssymb}
\usepackage{amsmath}

\usepackage[figuresright]{rotating}

\begin{document}

\begin{frontmatter}



\dochead{}

\title{Hybridization Of Electromagnetic, Spin and Acoustic Waves In Magnetic Having Conical Spiral Ferromagnetic Order}


\author[1]{Igor V. Bychkov}
\ead{bychkov@csu.ru}
\author[1]{Dmitry A. Kuzmin\corref{cor1}}
\ead{kuzminda89@gmail.com}
\author[2]{Vladimir G. Shavrov}

\address[1]{Chelyabinsk State University, 454001, Chelyabinsk, Br. Kashirinyh Street, 129, Russia}
\address[2]{The Institute of Radioengineering and Electronics of RAS, 125009, Moscow, Mokhovaya Street, 11-7, Russia}

\cortext[cor1]{Corresponding author}

\begin{abstract}
The spectrum of hybrid electromagnetic-spin-acoustic waves for magnetic having conical spiral ferromagnetic structure defined by heterogeneous exchange and relativistic interactions have been received. The possibility of resonant interaction of spin, electromagnetic and acoustic waves has been shown. The electromagnetic waves reflectance from the half-infinity layer of magnetic having conical spiral ferromagnetic order has been calculated for different values of external magnetic field (angle of spiral). The acoustic Faradey effect has been considered.
\end{abstract}


\end{frontmatter}


\section{Introduction}
\label{intro}
Recently, helicoidal (spiral) magnetic materials have attracted researchers' attention for their unusual physical properties \cite{1, 2}. The spiral magnetic structures contribute a number of features in the spectrum and dynamics of spin excitation in magnetic materials: band structure is observed, the nonreciprocity effect is manifested, i.e. difference between the velocity of wave transmission along and against the spiral axis. Previously, the spin-wave spectrum was calculated without taking into account of the effects of the electromagnetic retardation, and the electromagnetic wave spectrum was calculated without taking into account of the effects of the dynamic interaction of the electromagnetic field with the oscillations of the spins in the ferromagnetic spiral structure \cite{3, 4}. Earlier had been investigated the hybrid electromagnetic-spin, electromagnetic-spin-acoustic waves in the magnetic having simple spiral structure  \cite{5, 6}, and the hybrid electromagnetic-spin waves in the magnetic 
having conical spiral ferromagnetic structure, also termed "ferromagnetic spiral" \cite{7}. However the spectrum and dynamic properties of magnets in a phase “ferromagnetic spiral” are not studied enough. In the present work the spectrum of the hybrid electromagnetic-spin-acoustic waves in spiral magnetic structure of type “ferromagnetic spiral” is investigated. Also the reflection of electromagnetic waves from a surface of half-infinity magnetic material with a ferromagnetic spiral depending on the angle of spiral θ determined by an external magnetic field and Faradey effect are considered. Researches of spectrum of the coupled fluctuations in the modulated magnetic structures are spent in approach $L \gg a$, where $L$ = 2$\pi/q$ - the spiral period, $q$ - the wave number of spiral, $a$ - the lattice constant.


\section{The spectrum of hybrid spin, acoustic and electromagnetic waves}
\label{spectrum}
The ground state of a crystal is described by a vector of magnetization with components:
\begin{equation}\label{eq:ground}
M_{0x} = M_{0} \sin{\theta} \cos{qz}, M_{0y} = M_{0} \sin{\theta} \sin{qz}, M_{0z} = M_{0} \cos{\theta},
\end{equation}
where $M_{0}$ is magnetization of saturation, $q$ - wave number of a spiral, $\theta$ - an angle between a direction of magnetization and a spiral axis $z$. $\theta$ is defined by value of an external magnetic field. When $\theta = \pi/2$ the magnetic transforms from the phase of the “ferromagnetic spiral” into “simple spiral”, when $\theta$ = 0 - in the ferromagnetic phase.
	
The free energy of the crystal phase of “ferromagnetic spiral” has the form:
\begin{equation}\label{eq:free_energy}
F = \frac{\alpha}{2}\left({\frac{d \vec M}{dx_{i}}}\right)^2 + F_{in} + \frac{\beta_{1}}{2}{M_{z}}^2 + \frac{\beta_{2}}{2}{M_{z}}^4 - HM_{z} + 
b_{ijlm}M_{i}M_{j}u_{lm} + c_{ijlm}u_{ij}u_{lm},
\end{equation}
where $\vec M$ – the magnetization of the crystal; $u_{ij} = (\partial u_{i} / \partial x_{j} + \partial u_{j} / \partial x_{i})/2$ - the tensor of deformations; $\vec u$ - the displacement vector; $\alpha, \beta, b, c$ – the constants of inhomogeneous exchange, anisotropy, magnetostriction and elastic constant.
	
The term $F_{in}$, which causes inhomogeneous magnetization in the ground state for crystals with exchange spiral structure is:
\begin{equation}\label{eq:Fin_ex}
F_{in} = \frac{\gamma}{2}\left({\frac{d^{2} \vec M}{dx_{i}^{2}}}\right)^2,
\end{equation}
and for magnetics with a relativistic helicaloidal structure
\begin{equation}\label{eq:Fin_rel}
F_{in} = \alpha_{1} \vec M rot{\vec M},
\end{equation}
where $\gamma$ and $\alpha_{1}$ – constants of inhomogeneous exchange interaction and inhomogeneous relativistic interaction. In (\ref{eq:free_energy}) it is taken into account that the external magnetic field is directed along the axis of symmetry.
	
From the minimum of free energy with (\ref{eq:ground}) we obtain expressions for determining the angle $\theta$ through an external magnetic field $H$.
\begin{equation}\label{eq:H_theta}
H = M_{0} \cos{\theta} \left[\tilde{\beta}_{1} + h_{me} + \left(\tilde{\beta}_{2} + h_{me}/M_{0}^2\right) M_{0}^{2} \cos{\theta}^{2} +
\alpha q^{2} + \tilde{\Delta}\right],
\end{equation}
where ${\tilde \beta }_1$ and $ {\tilde \beta }_2 $ - the constants of anisotropy renormed by magnetostriction:
\begin{gather*}
{{\tilde \beta }_1} = {\beta _1} - \frac{{{c_{33}}{c_{11}} - c_{13}^2}}{{\Delta \left( {{c_{11}} - {c_{12}}} \right)}}{\left( {{b_{11}} - {b_{12}}} \right)^2}M_0^2 - \frac{{{c_{13}}}}{\Delta }\left( {{b_{33}} - {b_{31}}} \right)\left( {{b_{11}} - {b_{12}}} \right)M_0^2 + \\ 
+ \frac{{{c_{33}}}}{\Delta }\left( {{b_{13}} - {b_{12}}} \right)\left( {{b_{11}} - {b_{12}}} \right)M_0^2 + \frac{{b_{44}^2M_0^2}}{{2{c_{44}}}}, \\
{{\tilde \beta }_2} = {\beta _2} - \frac{{{c_{33}}{c_{11}} - c_{13}^2}}{{\Delta \left( {{c_{11}} - {c_{12}}} \right)}}{\left( {{b_{11}} - {b_{12}}} \right)^2} + \frac{{{c_{11}} - {c_{12}}}}{{2\Delta }}{\left( {{b_{33}} - {b_{31}}} \right)^2} + \frac{{{c_{33}}}}{\Delta }\left( {{b_{33}} - {b_{31}}} \right)\left( {{b_{11}} - {b_{12}}} \right) - \\ - \frac{{{c_{11}}}}{\Delta }\left( {{b_{13}} - {b_{12}}} \right)\left( {{b_{11}} - {b_{12}}} \right)
 - \frac{{b_{44}^2}}{{2{c_{44}}}} - \frac{{2{c_{13}}}}{\Delta }\left( {{b_{33}} - {b_{31}}} \right)\left( {{b_{13}} - {b_{12}}} \right), \\
 \Delta  = {c_{33}}\left( {{c_{11}} + {c_{12}}} \right) - 2c_{13}^2
\end{gather*}
	
For a spiral with the exchange interaction we have, $\gamma > 0$, $\alpha < 0$, $h_{me} = {\left(b_{11}-b_{12}\right)}^2 M_{0}^2 / \left( c_{11}-c_{12} \right)$, $q = {\left( -\alpha/2\gamma \right)}^{1/2}$, $\tilde{\Delta} = \gamma q^{4}$.
	
In the case of relativistic spiral, $\alpha_{1} \neq{0}$, $\alpha > 0$, $h_{me} = b^2 M_{0}^2/2\mu$, $q = \alpha_{1}/\alpha$, $\tilde{\Delta} = -2\alpha_{1} q$.
	
Note that the magnetoelastic coupling is not affected by the value of the wave number of the spiral $q$.
	
The tensor of the equilibrium deformations is:
\begin{gather} \label{eq:eq_def}
u_{xx}^0 = M_0^2\left( { - \frac{{{c_{33}}}}{{2\Delta }}\left( {{b_{11}} - {b_{12}}} \right){{\sin }^2}\theta  - \frac{1}{\Delta }\left[ {{c_{33}}\left( {{b_{13}} - {b_{12}}} \right) - {c_{13}}\left( {{b_{33}} - {b_{31}}} \right)} \right]{{\cos }^2}\theta } \right), \notag \\
u_{yy}^0 = u_{xx}^0,\quad u_{zz}^0 =  - \frac{{2{c_{13}}}}{{{c_{33}}}}u_{xx}^0 - \frac{1}{{{c_{33}}}}\left( {{b_{33}} - {b_{31}}} \right)M_0^2{\cos ^2}\theta ,\\
u_{xz}^0 =  - \frac{{{b_{44}}}}{{4{c_{44}}}}M_0^2\sin 2\theta \cos qz,\quad u_{yz}^0 =  - \frac{{{b_{44}}}}{{4{c_{44}}}}M_0^2\sin 2\theta \sin qz,\quad u_{xy}^0 = 0. \notag
\end{gather}
	
For solving a problem of getting the spectrum of hybrid waves one have to take into account the system of Maxwell's, Landau-Lifshitz and motion of an elastic medium equations: 
\begin{eqnarray}
\label{eq:system}
&{{\partial \vec M} / {\partial t}}  = g\left[ {\vec M{{\vec H}^{eff}}} \right],\quad {{\vec H}^{eff}} =  - {{\delta F}/{\delta \vec M}} \notag \\
&\rho \ddot u_{i} = {\partial {\sigma _{ik}}} / {\partial {x_k}},\quad {\sigma _{ik}} = {\partial F} / {\partial {u_{ik}}},\\
& rot\vec E =  - \dfrac{1}{c}\dfrac{\partial }{{\partial t}}\left( {\vec H + 4\pi \vec M} \right),\quad rot\vec H = \dfrac{\varepsilon }{c}\dfrac{{\partial \vec E}}{{\partial t}},\notag \\
& div\left( {\varepsilon \vec E} \right) = 0,\quad div\left( {\vec H + 4\pi \vec M} \right) = 0. \notag 
\end{eqnarray}
	
The linearized system of equations \eqref{eq:system} for Fourier components is:
\begin{eqnarray}
\label{eq:eq_system}
&\pm \omega {m_ \pm }\left( k \right) = \cos \theta \left[ {\omega _{2k}^ \pm  + \dfrac{1}{2}{\omega _{me4}}{{\sin }^2}\theta } \right]{m_ \pm }\left( k \right) + \dfrac{1}{2}{\omega _{me4}}{\sin ^2}\theta \cos \theta {m_ \pm }\left( {k \mp 2q} \right) - {\omega _{1k \pm q}}\sin \theta {m_z}(k \mp q) + \notag \\
&+ ig{b_{44}}M_0^2k\left[ {\dfrac{1}{2} - \dfrac{3}{2}{{\sin }^2}\theta } \right]{u_ \pm }\left( k \right) - \dfrac{i}{2}g{M_0}^2{b_{44}}{\sin ^2}\theta \left( {k \pm 2q} \right){u_ \pm }\left( {k \mp 2q} \right) - \notag \\
&- ig\left( {{b_{33}} - {b_{31}}} \right)M_0^2\sin 2\theta \left( {k \pm q} \right){u_z}\left( {k \mp q} \right) + g{M_0}\sin \theta {h_z}\left( {k \mp q} \right) - g{M_0}\cos \theta {h_ \pm }\left( k \right), \notag \\ 
&\omega {m_z}\left( k \right) = \dfrac{1}{2}\sin \theta \left[ {\omega _{2k - q}^ - {m_ - }\left( {k - q} \right) - \omega _{2k + q}^ + {m_ + }\left( {k + q} \right)} \right] + \dfrac{1}{2}g{M_0}\sin \theta \left[ {{h_ + }\left( {k + q} \right) - {h_ - }\left( {k - q} \right)} \right] - \\
&- \dfrac{i}{4}g{b_{44}}M_0^2\sin 2\theta \left[ {\left( {k - q} \right){u_ - }\left( {k - q} \right) - \left( {k + q} \right){u_ + }\left( {k + q} \right)} \right],\notag \\
&\left[ {{\omega ^2} - s_t^2{k^2}} \right]{u_ \pm }\left( k \right) = \dfrac{i}{\rho }k{b_{44}}{M_0}\left[ {\sin \theta {m_z}\left( {k \mp q} \right) + \cos \theta {m_ \pm }\left( k \right)} \right], \notag \\
&\left[ {{\omega ^2} - s_l^2{k^2}} \right]{u_z}\left( k \right) =  - 2i\left( {{b_{33}} - {b_{31}}} \right)k{M_0}\cos \theta {m_z}{\left( k \right)} / \rho , \notag \\
&\left[ {{\omega ^2} - {k^2}{v^2}} \right]{h_ \pm }\left( k \right) =  - {\omega ^2}4\pi {m_ \pm }\left( k \right),\quad {h_z}\left( k \right) =  - 4\pi {m_z}\left( k \right). \notag
\end{eqnarray}
	
Here, we introduce the following notation: $v = c / \sqrt{\varepsilon}$ - velocity of propagation of electromagnetic waves in a magnetic, $\varepsilon$ - dielectric constant, $s_t = \sqrt{c_{44} / \rho}, \ s_l = \sqrt{c_{33} / \rho}$ - velocity of propagation of transversal and longitudinal acoustic waves, respectively,
\begin{eqnarray}
\label{eq:notation}
& \omega _{1k} = \omega_{10} + g M_0 \sin^2{\theta} L_{\parallel}(k), \ \omega _{2k}^\pm = \omega_{20} + g M_0 L_{\bot}(k), \notag \\
& L_\bot ^ \pm (k) =  - \alpha \left( {q^2} - {k^2} \right) - \gamma \left( {q^4} - {k^4} \right) + 2{\alpha _1} \left( q \mp k \right), \notag \\
& {L_{\parallel}}(k) =  - \alpha \left( {q^2} - {k^2} \right) - \gamma \left( {q^4} - {k^4} \right) + 2{\alpha _1} q, \\
& \omega_{me4} = g M_0 h_{me4} = g b_{44}^2 M_0^3 / c_{44}, \ \omega_{20} = \omega_{me4} \cos^2{\theta} , \notag \\
& \omega_{10} = g M_0 \left[ h_{me4} - \sin^2{\theta} \left( \tilde {\beta_1} + M_0^2 \cos^2{\theta} \left( \tilde{\beta_2} + 2 \beta_2 \right) + h_{me} \sin^2{\theta} \right) \right] . \notag
\end{eqnarray}

In system of equations (\ref{eq:eq_system}) we have to add the condition of constancy of the modulus of the magnetization $\left| {\vec M} \right| = const$, what for the Fourier components of the magnetization is:
\begin{equation}
\label{eq:cond_of_const}
\sin{\theta} \left[ {{m_ - }\left( {k - q} \right) + {m_ + }\left( {k + q} \right)} \right] + 2 {m_z}\left( k \right) \cos{\theta} = 0.
\end{equation}
	
Using the ordinary values of the constants for magnetic with exchange spiral (TbMn$_{2}$O$_{5}$) $g = 2\cdot10^7 s^{-1} erg^{-1}$, $M_0 \sim 10^3 Oe$, $\alpha_{1} \sim 10^{-28} cm^4$, $\alpha \sim -10^{-14} cm^2$, $q \sim 10^7 cm^{-1}$,  and in the case of relativistic spiral (CsCuCl$_{3}$) $\alpha \sim 10^{-12} cm^2$, $\beta \sim 1$, $a \sim 10^{-8} cm$, $q \sim 10^4 cm^{-1}$,  from equations \eqref{eq:eq_system}, we obtain the spectrum of coupled electromagnetic-spin-acoustic waves.
	
\begin{figure}[h!]
\center{\includegraphics[width=90mm]{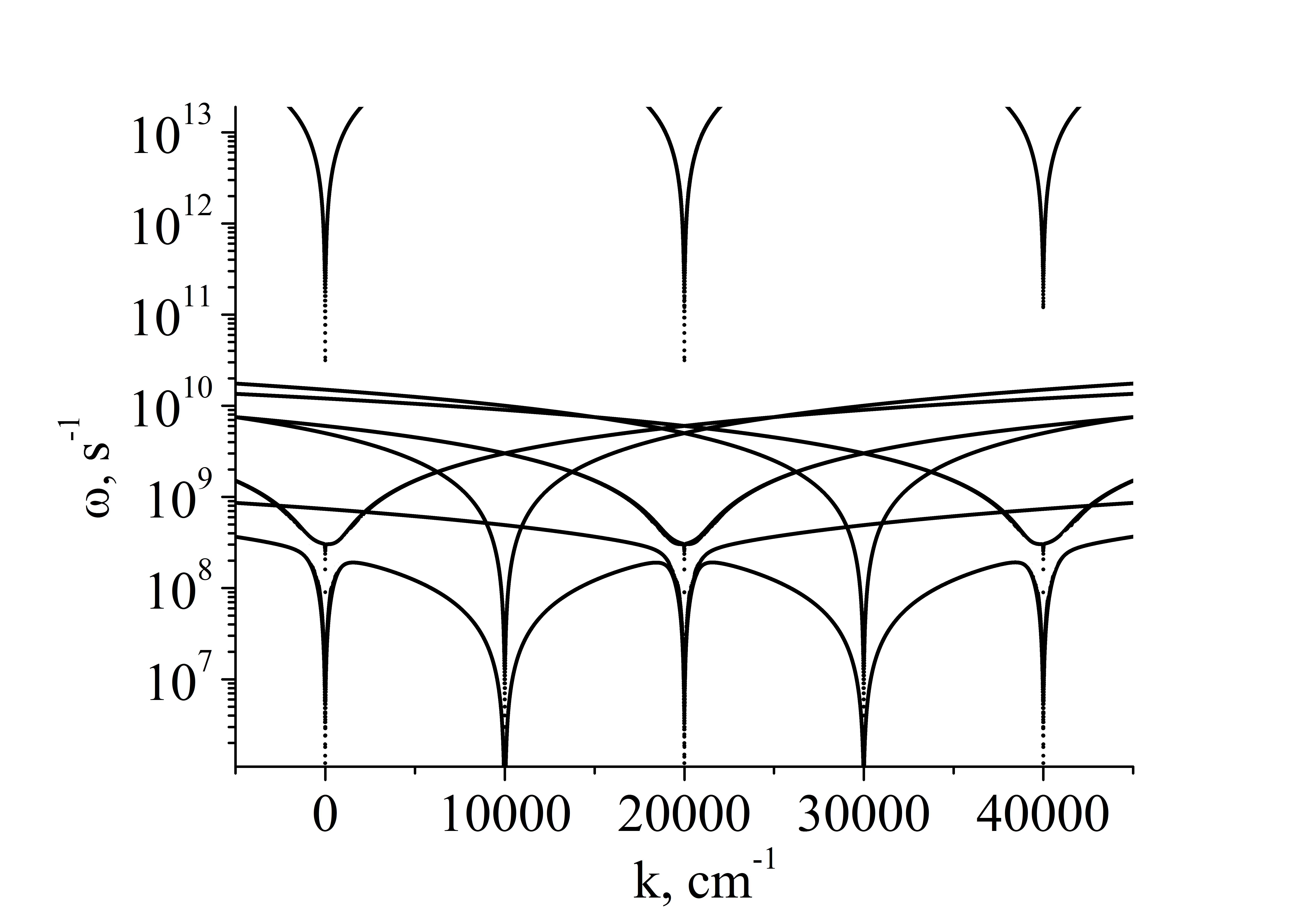}}
\caption{Spectrum of coupled oscillations for $\theta = \pi/4$.}
\label{ris:spectrum}
\end{figure}

Changing $\thetaθ$ the range $0 \leq \theta \leq \pi/2$ , we can calculate the spectrum for the ferromagnetic spiral. Figure \ref{ris:spectrum} shows the dependence $\omega (k)$ for $\theta = \pi/4$ in the case of relativistic spiral.
	
\begin{figure} [h!]
\center{\includegraphics[width=90mm]{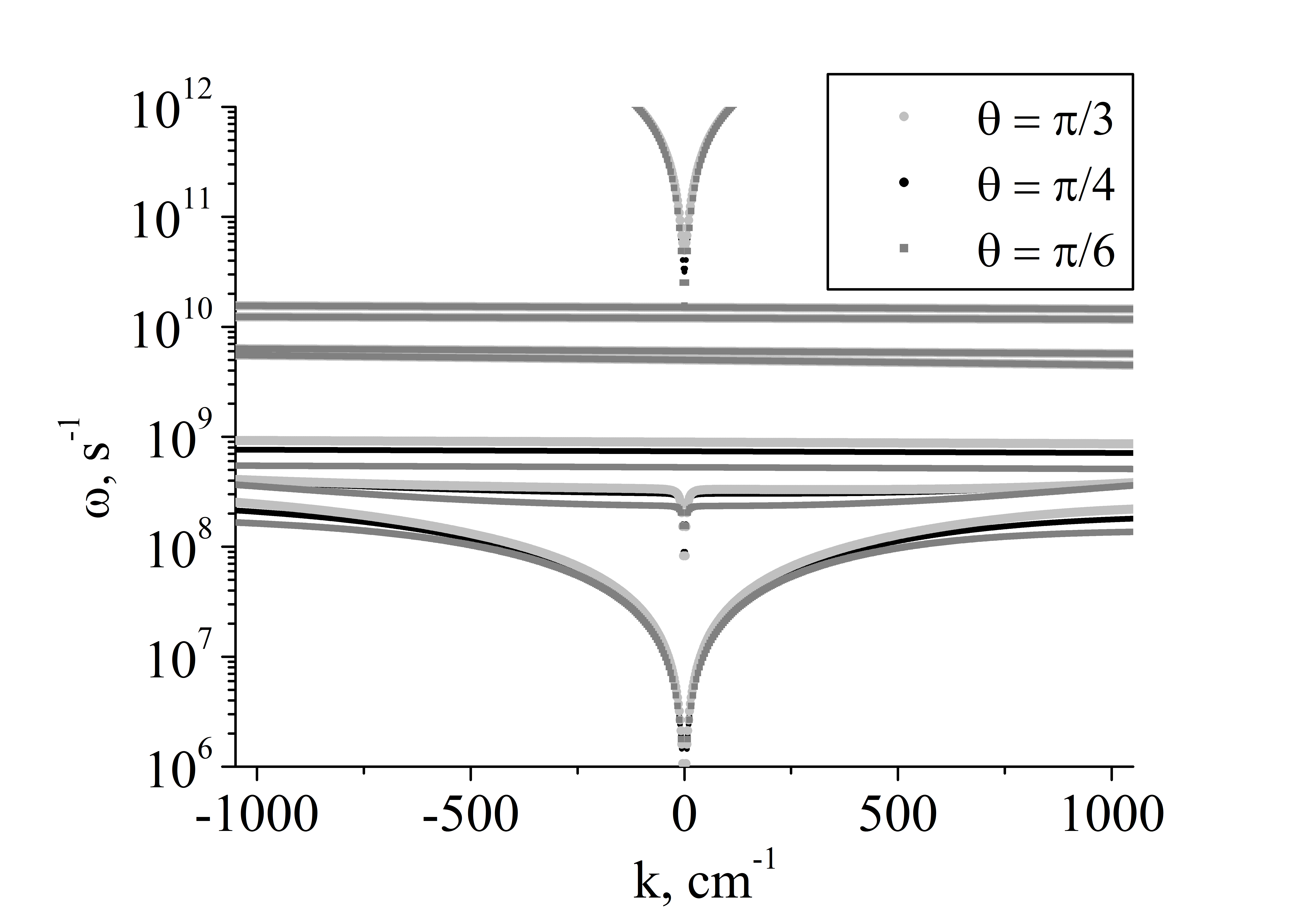}}
\caption{Spectrum of coupled oscillations for different $\theta$ near $k = 0$.}
\label{ris:spectrum_0}
\end{figure}

To get more details and study the dynamic of changing band gaps width with changing external magnetic field value  let us consider the dependence $\omega (k)$ for different values of $\theta$ near $k = 0$ (Fig. \ref{ris:spectrum_0}).
	
It is seen that all spectra have a band structure. At certain frequencies the gap (window opacity) is observed as for electromagnetic such for acoustic waves. These band gaps appear due to the resonant interaction of spin, acoustic and electromagnetic waves in a magnet. From Fig. \ref{ris:spectrum_0} we can see that with decreasing angle the electromagnetic band shifts toward lower frequencies and its width decreases. Calculations show that in the case of exchange spiral, a band of opacity is much narrower than in the case of relativistic one. Note also that the magnitude of the interaction of spin, acoustic and electromagnetic waves depends on the angle $\theta$.

\section{The reflection of electromagnetic waves from a magnetic having conical spiral ferromagnetic order}
\label{reflection}
Let us now investigate the reflection of electromagnetic waves from a half-infinity layer of magnetic material in phase of “ferromagnetic spiral”. Consider the normal incidence of electromagnetic wave $ k \parallel q \parallel z $. Restrict ourselves to small wave numbers $k \ll q$. The system of boundary conditions include the continuity of normal components of the magnetic and electric fields, the tangential component of the electric and magnetic fields, the vanishing of the derivative of the magnetization and the absence of stress at the boundaries of the magnet:  $H_\tau ^{(e)} = H_\tau ^{(i)}, \ E_\tau ^{(e)} = E_\tau ^{(i)}, \ B_n^{(e)} = B_n^{(i)}, \ D_n^{(e)} = D_n^{(i)},\ \dfrac{{\partial \vec m}}{{\partial {x_k}}}{n_k} = 0, \ \sigma _{jk}^{(i)}{n_k} = 0$. Indexes $(i)$ and $(e)$ denotes quantities inside and outside the magnet, respectively, $\vec n$ - normal to the surface. Taking into account the number of roots of dispersion equation in this approximation, the system of boundary conditions in 
the 
cyclical components of the magnetic field becomes:
\begin{eqnarray} \label{eq:bound_cond}
&{h_{0 \pm }} + {h_{R \pm }} = \sum\limits_{j = 1}^3 {{h_{j \pm }}} ; \quad {h_{0 \pm }} - {h_{R \pm }} = \sum\limits_{j = 1}^3 {\dfrac{{{k_{j \pm }}}}{{{k_0}\varepsilon }}{h_{j \pm }};} \quad \sum\limits_{j = 1}^3 {{h_{jz}} = 0;} \notag \\
&\sum\limits_{j = 1}^3 {{k_{j \pm }}} \left( {{\omega ^2} - {v^2}{k_{j \pm }}^2} \right){h_{j \pm }} = 0; \ \quad \sum\limits_{j = 1}^3 {{k_{j \pm }}{h_{jz}}}  = 0;\\
&\cos \theta \sum\limits_{j = 1}^3 {\dfrac{{{\omega ^2} - {v^2}{k_{j \pm }}^2}}{{{\omega ^2} - {s_t}^2{k_{j \pm }}^2}}} {h_{j \pm }} + \sin \theta \sum\limits_{j = 1}^3 {\dfrac{{{\omega ^2}}}{{{\omega ^2} - {s_t}^2{k_{j \pm }}^2}}} {h_{jz}} = 0; \quad \sum\limits_{j = 1}^3 {\dfrac{{{s_l}^2{k_{j \pm }}^2}}{{{\omega ^2} - {s_l}^2{k_{j \pm }}^2}}} {h_{jz}} = 0. \notag
\end{eqnarray}

Fields  define wave reflected from the surface of the magnetic, $ k_{i \pm} $  - solutions of the dispersion equation. Solving (\ref{eq:bound_cond}) with (\ref{eq:eq_system}), we find the reflection coefficient of electromagnetic waves $ R_ \pm  = \left| h_{R \pm }/h_{0 \pm } \right|^2$ . 
\begin{figure} [h!]
\center{\includegraphics[width=90mm]{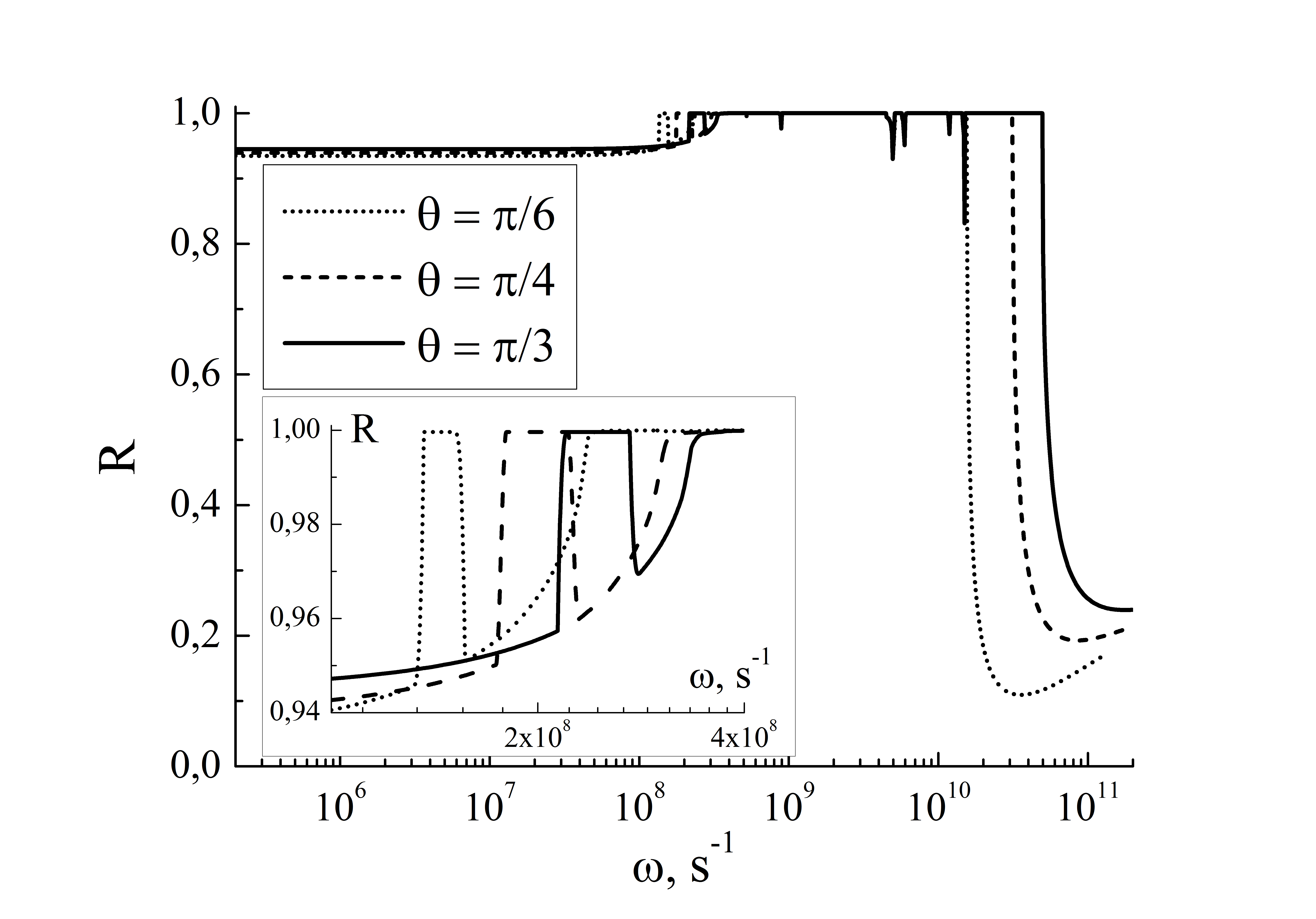}}
\caption{Frequency dependances of reflectance for different $\theta$.}
\label{ris:reflection}
\end{figure}
	
Figure \ref{ris:reflection} shows the frequency dependence of the reflection coefficient of electromagnetic waves from an interface of half-infinity magnetic material with a ferromagnetic spiral at different angles of spiral.
From the frequency dependence for different angles $\theta$ can be seen that with increasing angle $\theta$ of ferromagnetic spiral (and thus decrease the external magnetic field) increases the band gap (window opacity) and shift it to higher frequencies. Also there is a peak at frequencies of resonant spin-acoustic interaction, which shifts to higher frequencies with increasing $\theta$.

\section{Acoustic Faradey effect}
\label{faradey}
\begin{figure} [h!]
\center{\includegraphics[width=90mm]{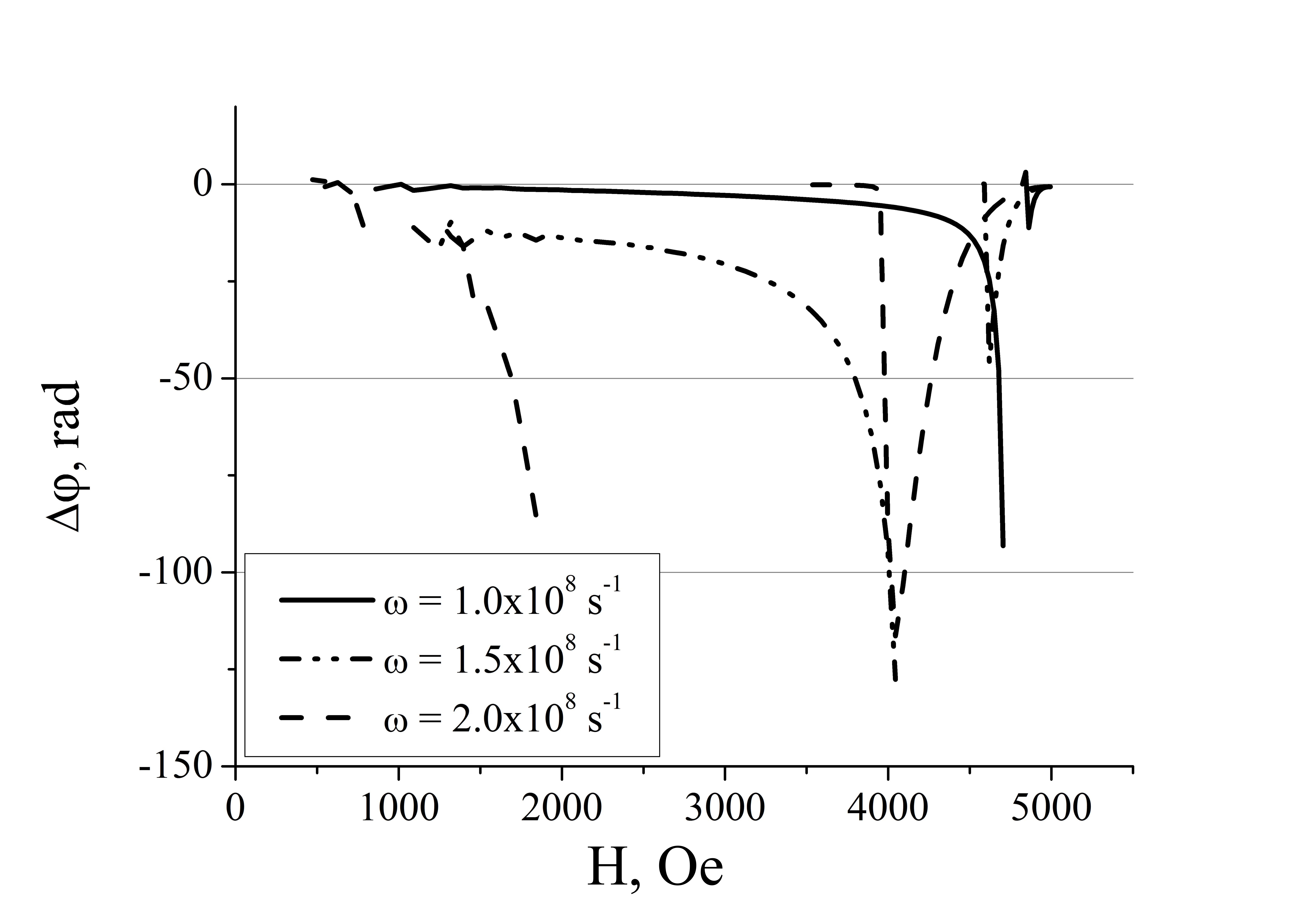}}
\caption{The dependence of the angle of rotation of the polarization plane from the external magnetic field $H$.}
\label{ris:faradey}
\end{figure}

Let us now consider the acoustic Faraday effect, i.e. dependence of the angle of rotation of the polarization plane of acoustic wave from the external magnetic field. Let the magnet in a phase of "ferromagnetic spiral" falls linearly polarized acoustic wave. It can be represented as a superposition of two acoustic waves of different circular polarization. From (\ref{eq:eq_system}) for waves of different polarizations obtain different wave vectors.
Since the wave vectors of the waves are different, so the refractive indices also will be different, therefore, will be observed the rotation of the polarization plane by an amount $\Delta \varphi  = \Delta kl$ , where $\Delta k = \left( {k_ + } - {k_ - } \right) / 2$ . Based on the solutions (\ref{eq:eq_system}), it depends on the frequency of the incident acoustic wave and reaches maximum values near the acoustic band gap. The dependence of the angle of rotation of the polarization plane from the external magnetic field for magnetic layer thickness of 1 cm is shown in Figure \ref{ris:faradey}.
\section{Conclusions}
\label{conclusion}
Studies on the hybrid electromagnetic-spin-acoustic waves in magnetic materials with a helical magnetic structure, defined by the inhomogeneous exchange and relativistic interactions in the phase of “ferromagnetic spiral”, showed that the spectrum of the coupled waves has band structure. The band gap depends on the angle of the ferromagnetic spiral, and hence on the external magnetic field. Increase in the angle (decrease in the magnetic field) leads to an increase in band gap, the maximum gap (window opacity) is observed at $\theta = \pi/2$ , i.e. at the phase transition “ferromagnetic spiral” – “simple spiral”. The possibility of resonant interaction of spin, acoustic and electromagnetic waves has been shown. The value of the interaction of the waves depends on $\theta$. The frequency dependence of the reflection coefficient of electromagnetic waves from the plate of the magnetic with a ferromagnetic spiral at different angles of the helix has been calculated. As the angle increases the opacity region 
broadens and shifts to higher frequencies. 
The angle of rotation of the polarization increases near the band gap. There is a peak of rotation of the polarization plane, which shifts to lower fields if the frequency increases





\end{document}